# Two-dimensional 5*d* multiferroic $W_3Cl_8$: breathing Kagome lattice and tunable magneto-optical Kerr effect

Di Hu, Haoshen Ye, Ning Ding, Kaidi Xu, Shan-Shan Wang, Shuai Dong,[*] and Xiaoyan Yao[†]

*Key Laboratory of Quantum Materials and Devices of Ministry of Education,*

*School of Physics, Southeast University, Nanjing 211189, China*

[†]Email: yaoxiaoyan@seu.edu.cn   [*]Email: sdong@seu.edu.cn

**Abstract:** Owing to the strong spin-orbit coupling and the related fascinating physical properties, heavy 5*d* transition-metals exhibit desirable application prospects. However, up to now, the 5*d* magnetic materials are still very limited, especially very rare for tungsten. In this work, we theoretically predict a two-dimensional multiferroic $W_3Cl_8$ monolayer. Intrinsic 5*d* magnetism of tungsten is activated by the W ions' fractional valence in a breathing Kagome lattice of reduced effective dimension. A coplanar Y-type antiferromagnetism composed by ferromagnetic $W_3$ trimers is confirmed as the magnetic ground state. The spontaneous ferroelectric polarization mainly originates from the ion displacement induced by the breathing distortion of Kagome lattice. An intrinsic magneto-optical Kerr effect with sizable Kerr angle can be observed to detect this trimeric Y-type antiferromagnetism, and it depends strongly on the detailed magnetic order. Thereby, we propose a general scheme for realizing more 5*d* magnetism in two-dimensional multiferroic systems.

# Introduction

Compared to the well-known $3d$ magnetism and substantial $3d$ magnetic materials, the intrinsic magnetism from $5d$ electrons was relatively much less reported. The ferromagnetic order can be stabilized in $3d$ metal bulks, but it hardly exists in $5d$ metals at room temperature due to the wider bands and weaker on-site Coulomb interaction [1]. In 1992, ferromagnetism of $5d$ transition metal monolayer was predicted for the first time [2]. In the following decades, it was much highlighted that the unconventional bond-dependent Kitaev magnetic interaction was believed to be substantial in the hexagonal iridates [3,4]. In addition, $5d$ magnetism of Os and Re had been widely investigated in double perovskites [5,6]. However, up to now, the $5d$ magnetic materials are still very limited, especially for tungsten only very few were reported [7,8]. Owing to the strong spin-orbit coupling, heavy $5d$ transition-metals exhibit fascinating physical properties and desirable application prospect, such as Kitaev magnetism, topologically nontrivial states, spintronic technology and quantum computing, especially when combined with magnetism [9-11].

Generally, the reduction of dimension can suppress band width and thus help to electron localization. However, it is not easy to achieve $5d$ magnetism of tungsten even for low-dimensional structures, because the $5d$ orbitals are more delocalized than those of $3d$ elements. As a pioneering work, Xu *et al*. predicted a multiferroic $ReWCl_6$ monolayer with magnetic moment of 1 $\mu_B$ coming from $W^{5+}$-$d^1$ electrons, and investigated the electrical control of magnetic phase transition [12]. Note that the crystal field originating from halide ions is weaker than that of chalcogen, and a breathing Kagome lattice with less bonds between sites may present a lower effective dimension, approaching an isolate system. Two-dimensional (2D) Kagome halides might provide a better way to trigger the $5d$ magnetism of tungsten. Since the 1960s, layered $Nb_3X_8$ ($X$=Cl, Br, I) bulks with a breathing Kagome lattice had been synthesized and investigated [13-15]. In 2019 and 2022, the $Nb_3I_8$ and $Nb_3Cl_8$ monolayers were exploited from their bulks experimentally [16,17]. The breathing Kagome lattice remains down to the monolayer limitation. Recently, this family of 2D Kagome halides is attracting growing interest due to the diverse intriguing phenomena, such as topological flat bands [17,18], intrinsic ferrovalley [19] and multiferroicity [20,21].

In this work, based on first-principles calculations and atomistic simulation, we predict a 2D multiferroic $W_3Cl_8$ monolayer. The intrinsic $5d$ magnetism is triggered on W due to its fractional valence in a breathing Kagome lattice of reduced effective dimension. A coplanar trimeric Y-type antiferromagnetism is confirmed as magnetic ground state. Meanwhile, the out-of-plane ferroelectric polarization mainly originates from the ion displacement induced by the breathing distortion. Moreover, an intrinsic magneto-optical Kerr effect (MOKE) with sizable Kerr angle is revealed in this antiferromagnetic (AFM) monolayer, which shows different dependences on detailed magnetic order for different chiralities. Thus, we propose a general scheme for realizing 5$d$ magnetism of tungsten in 2D multiferroic systems, and provides an excellent platform for exploring magneto-optical phenomena and new technological applications.

**Methods**

The spin-polarized calculation is implemented in the framework of density functional theory (DFT) [22], as performed using the Vienna *ab initio* Simulation Package (VASP). The exchange-correlation potential is characterized by the Perdew-Burke-Ernzerhof (PBE) of the generalized gradient approximation (GGA) [23]. The cutoff energy is set to 500 eV. The convergent criterion of energy is set as $10^{-5}$ eV, and that of the Hellman-Feynman forces is 0.01 eV/Å for all the calculations. A Γ-centered $k$-grid of 5×5×1 is used to sample the Brillouin zone. A vacuum region of 30 Å is set to avoid interaction among the adjacent layers along the $c$-axis.

Since the Hubbard $U$ for 5$d$ element is relatively small and it is sometimes ignored in previous works [24-26], we have tested the DFT+$U$ method with effective $U$ ($U_{eff}$) for the 5$d$ orbitals of W. The breathing Kagome lattice and magnetic ground state are robust for different $U_{eff}$, and the main results of ferroelectricity and MOKE are not significantly influenced (see Note 1 in the Supplemental Material [27] for details). Therefore, subsequent calculations are performed without $U$.

Based on the density functional perturbation theory (DFPT) [28], the phonon spectrum is calculated on a 2×2×1 supercell by using the PHONOPY code [29], where a rotational sum rule is enforced on the force constants as implemented in the Hiphive package [30]. The ferroelectric polarization is evaluated by the standard Berry phase method [31,32].

To investigate the magnetic ground state of W$_3$Cl$_8$ monolayer, an atomistic simulation has been performed with the Landau-Lifshitz-Gilbert (LLG) equation as executed using the SPIRIT package [33]. A supercell of 70×70×1 sites with periodic boundary conditions and 2×10$^5$ iterations are used to achieve a stable state at zero temperature.

For 2D finite thick materials with threefold rotational symmetry, the complex Kerr angle can be evaluated as [34]:

$$\theta_K + i\eta_K = \frac{2\xi_{xy}}{1-(\xi_{xx}+n_s)^2-\xi_{xy}^2}, \tag{1}$$

where $\theta_K$ specifies the Kerr rotation angle and $\eta_K$ specifies the ellipticity. $\xi_{xy}=\sigma_{xy}Z_0$ and $\xi_{xx}=\sigma_{xx}Z_0$, where $Z_0$ is the vacuum impedance. $n_s$ is the vacuum refractive index. $\sigma_{xy}$ and $\sigma_{xx}$ are elements of the optical conductivity tensor (OCT) which can be obtained via the first principles calculations [34]. The magnetic point group (MPG) is analyzed using FINDSYM [35]. During the calculation of MOKE and magnetic anisotropy energy (MAE), the spin-orbit coupling (SOC) is enabled.

## Results and discussion

### A. Structure: trimerization

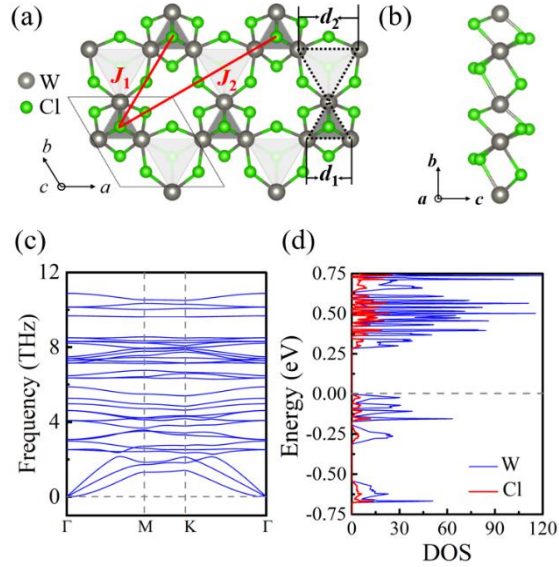

**FIG. 1.** (a) Crystal structure of $W_3Cl_8$ monolayer. $W_3$ trimers are denoted by the dark gray triangles. $d_1$ and $d_2$ represent the intra-trimer and inter-trimer W-W distance. $J_1$, and $J_2$ denote the exchange couplings between the nearest-neighboring and second-neighboring trimers. (b) Side view. (c) Phonon spectrum. (d) Atom-projected density of states (DOS) with SOC considered.

Our structural optimization finds that $W_3Cl_8$ monolayer has a polar space group $P3m1$ (No. 156), and the optimized lattice constant is 6.670 Å. W ions are sandwiched between the upper and lower Cl layers, as shown in Fig. 1(a-b). The distorted $WCl_6$ cages are connected in the manner of edge-sharing mode, forming a breathing Kagome lattice. As a result, half W triangles shrink into trimeric clusters (i.e. $W_3$ trimers). There are two distinct W-W distances:

the intra-trimer $d_1$=2.754 Å and the inter-trimer $d_2$=3.916 Å. The W$_3$ trimers form a triangular lattice.

According to the phonon spectrum plotted in Fig. 1(c), W$_3$Cl$_8$ monolayer with breathing Kagome lattice is dynamic stable, since there is no imaginary frequency within the numerical precision. In addition, the thermodynamical stability is also evaluated by estimating the average cohesive energy ($E_{coh}$):

$$E_{coh} = (3E_W + 8E_{Cl} - E_{total})/11, \qquad (2)$$

where $E_{total}$ is the energy of monolayer W$_3$Cl$_8$. $E_W$ and $E_{Cl}$ are the energies of single W and Cl atoms, respectively. The calculated $E_{coh}$ is 3.96 eV/atom, which is comparable to those of other common monolayers like Be$_2$C (4.86 eV/atom) [36] and Cu$_2$Si (3.46 eV/atom) [37]. These calculations confirm the stability of W$_3$Cl$_8$ monolayer.

As shown in Fig. 1(d), the W$_3$Cl$_8$ monolayer is a semiconductor with a small band gap ~0.3 eV (also see Fig. S2 in Supplemental Material [27]). Such a gap is opened by the breathing mode of Kagome lattice, as a characteristic of Peierls phase transition, since the system is metallic without the breathing mode (see Fig. S3 in Supplemental Material [27]). Such a semiconducting behavior is general for this material family with fractional valence and breathing Kagome lattice [18-20,38].

## B. Magnetism: noncollinear Y-AFM order

Under the octahedral crystal field, the $d$ orbitals are always split into two groups: the low-lying $t_{2g}$ triplets and energy higher $e_g$ doublets. The trimerization of W ions leads to the reconstruction of molecular orbitals and further splitting of these $t_{2g}$ orbitals into 1$e$, 1$a_1$, 2$e$, 2$a_1$, 3$e$, and 1$a_2$ [19,39]. The three W ions in each trimer share ten electrons, considering the +8 valence of W$_3$ trimer. For 5$d$ orbitals with weak Hund coupling, a low spin state is expected. Even though, as shown in Fig. 2(a), a local magnetic moment of 2.00 $\mu_B$ per trimer remains expectable thanks to the double-degenerated 2$e$ level. Indeed, our DFT calculation confirms this 2.00 $\mu_B$ local moment per trimer, despite the value of $U_{eff}$ used in the calculation. As a cluster magnet, this magnetic moment is delocalized on the trimer, which can not be assigned to each individual W ion. Such a robust cluster magnetism is conventional for the materials with a breathing Kagome lattice, such as Nb$_3$X$_8$, Ti$_3$X$_8$ (X = Cl, Br, or I) [19,20] and $A$Mo$_3$O$_8$ ($A$ represents the interstitial cations) [40,41].

Then the MAE of this moment is calculated by setting a ferromagnetic (FM) order. As shown in Fig. 2(b), the energy for spin pointing along the $z$-axis is 22.3 meV per trimer higher than that of spin in the *xy*-plane, implying a strong magnetic anisotropy of easy-plane type. Thus, in the following, we only consider those textures with spins lying in the *xy*-plane.

Before the determination of the magnetic ground state, we first check the magnetic texture within $W_3$ trimer, since the intra-trimer magnetic coupling should be much stronger than the inter-trimer ones. Three most possible candidates for the W triangle are the FM, ferrimagnetic (FiM), and noncollinear Y-type antiferromagnetic (Y-AFM) states, as illustrated in Fig. 2(c). Our DFT calculation confirms that the FM order possesses the lowest energy for each $W_3$ trimer, as shown in Table S3 in Supplemental Material [27].

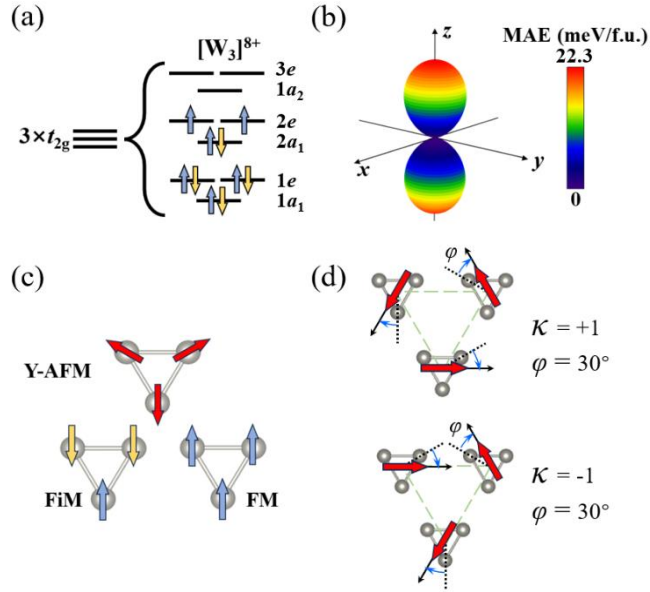

**FIG. 2.** (a) Sketch of the *d* orbitals splitting for $W_3$ trimer and corresponding electron filling. (b) Angular dependence of calculated MAE on spin orientation. The *z*-axis is along the [001] crystallographic orientation. (c) Three possible magnetic textures within one $W_3$ trimer, where the arrow shows the magnetic moment orientation of each W ion. Among them, the FM one has the lowest energy. (d) The trimeric Y-AFM order of $W_3$-spin lattice, where the arrow shows the magnetic moment orientation of each FM $W_3$ trimer. The chiralities ($\kappa$=+1 and $\kappa$=-1) are distinguished. $\kappa$ is calculated for the downward triangle in the counterclockwise sequence. The azimuthal angle $\varphi$ is defined to measure the clockwise synchronous rotation of all spins, and the dotted lines mark the location of $\varphi$=0.

By treating the magnetic moment of each FM $W_3$ trimer as a unit ($W_3$-spin), the magnetic order of $W_3$-spin lattice is calculated by comparing the energies of several possible states, as shown in Fig. 2(d) and Fig. S4 in Supplemental Material [27]. It is found that the trimeric Y-AFM order with 120º angles between $W_3$-spins owns the lowest energy (Table S4 in Supplemental Material [27]). Then the magnetic coefficients $J_1$ and $J_2$ between the nearest-neighboring and second-neighboring $W_3$-spins (as depicted in Fig. 1(a)) can be obtained by comparing the DFT energies of different orders (see Note 3 in Supplemental Material [27] for details). With normalized $W_3$-spin, the obtained $J_1$ is ~2.2 meV and $J_2$ is ~0.3 meV. Such a small $J_2$ is reasonable considering the long distance between the second-neighboring $W_3$-spins. Based on these couplings, the atomistic simulation is performed. The result indicates that the noncollinear Y-AFM state composed by FM $W_3$ trimers is indeed the ground state for the $W_3$-spin lattice.

This Y-AFM order of $W_3$-spins on triangular lattice has diverse morphologies. For example, the vector chirality ($\kappa$) can be defined as [42]:

$$\kappa = \frac{2}{3\sqrt{3}} \sum_{<ij>} [S_i \times S_j]_z, \tag{3}$$

where $<ij>$ refers to the nearest-neighboring $W_3$-spins. As indicated in Fig. 2(d), $\kappa$ can be calculated for the downward triangle in a counterclockwise sequence. For the in-plane Y-AFM state, $\kappa=+1$ and $-1$ just represent the two opposite chiralities. In addition, the trimeric Y-AFM state may have different angles respect to the crystal lattice. Here, the azimuthal angle $\varphi$ is used to measure the clockwise synchronous rotation of all spins, as illustrated in Fig. 2(d).

## C. Ferroelectricity

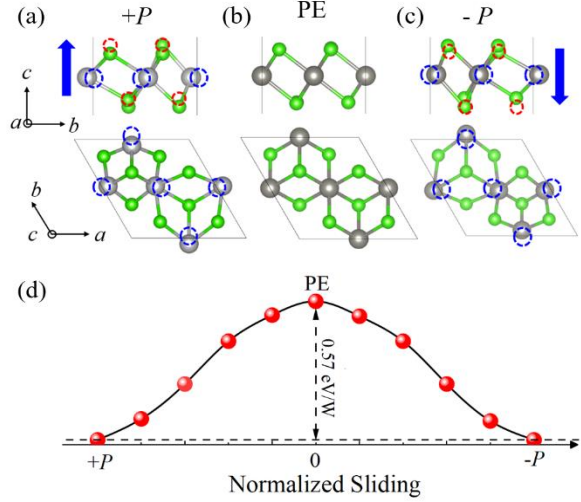

**FIG. 3.** Side and top views of the $W_3Cl_8$ monolayer for (a) ferroelectric $+P$ phase, (b) paraelectric (PE) phase and (c) ferroelectric $-P$ phase. Red and blue dashed circles indicate the positions of Cl and W ions in the PE phase, respectively. (d) The CI-NEB result of the transition path from the $+P$ state to $-P$ state via the intermediate PE phase.

As mentioned above, the semiconducting properties of $W_3Cl_8$ monolayer is stabilized by the breathing distortion of Kagome lattice, which can also generate a vertical ferroelectric polarization. As illustrated in Fig. 3(a), the in-plane shrinking of each $W_3$ trimer squeeze the middle Cl ions along the $c$ axis, which induces a polarization. $+P$ and $-P$ states correspond to different phases of $W_3$ trimerization. The standard Berry phase calculation indicates that the total polarization of $W_3Cl_8$ monolayer reaches 0.394 $\mu C/cm^2$ (with $\kappa=+1$ in $+P$ phase), if the height of $W_3Cl_8$ (3.467 Å) is taken as the thickness.

In fact, the ferroelectric polarization also contains a part of contribution from the noncollinear $W_3$-spin texture via the SOC. Such a magnetism-induced ferroelectric polarization ($P_m$) is determined by its vector chirality [43,44], and its direction should be normal to the crystal plane. By comparing the net polarizations of $\kappa=+1$ and $\kappa=-1$ configurations, $P_m$ is estimated as 0.005 $\mu C/cm^2$, only 1.27% of total polarization. Thus, the magnetic contribution to polarization is negligible, and $W_3Cl_8$ monolayer can be considered as a typical type-I multiferroicity [45].

Such a ferroelectric polarization can be flipped between the $+P$ and $-P$ states by a breathing process of the Kagome network. The intermediate state without $W_3$ trimerization can be considered as the parent paraelectric (PE) phase, as illustrated in Fig. 3(b). Then the ferroelectric switching shown in Fig. 3(a-c) can be simulated by the climbing image nudged elastic band (CI-NEB) method [46], which leads to an energy barrier shown in Fig. 3(d). Note here the energy barrier can only be considered as a theoretical upper limit, while the real barrier may be lower via more complex switching paths. Anyway, due to the ultrathin thickness and the vertical polarization of $W_3Cl_8$ monolayer, a sufficiently strong electric field can be applied perpendicularly to achieve ferroelectric switching [20,47].

According to the DFT calculations, all the trimeric Y-AFM states with different $\varphi$ and the same $\kappa$ are nearly degenerate in energy, and the energy of $\kappa = +1$ is about 1 meV per unit cell lower than that of $\kappa = -1$ in the same $+P$ phase, due to the different directions of $P_m$.

**D. Magneto-optical Kerr effect**

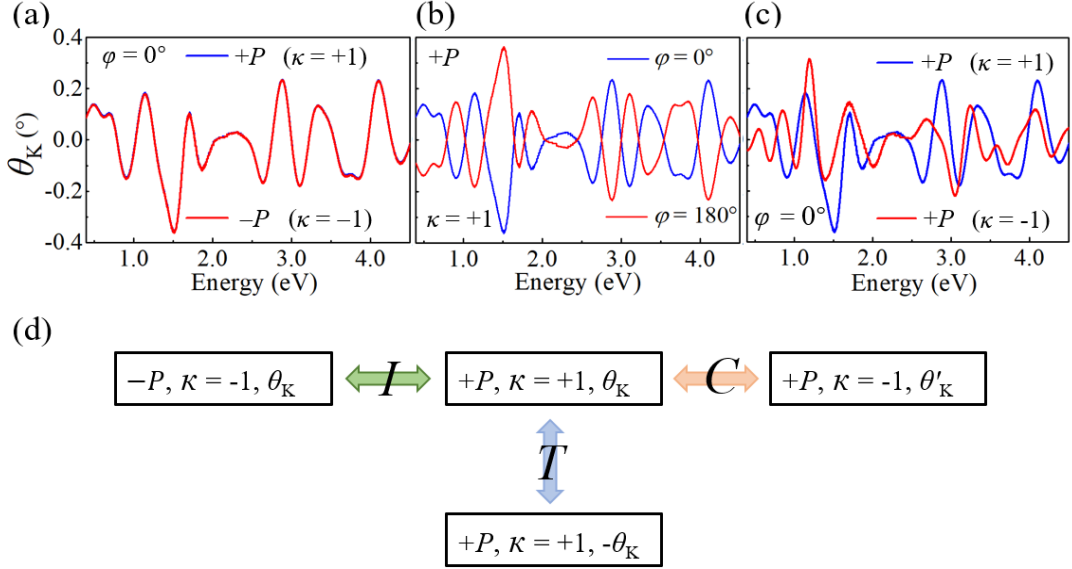

**FIG. 4.** Kerr spectra of $W_3Cl_8$ monolayer under (a) spatial inversion ($I$) operation and (b) time reversal ($T$) operation. (c) The Kerr spectra in $+P$ phase for different $\kappa$. (d) The symmetry relations under different operations, namely $I$, $T$, and the pure chirality switch ($C$) between $\kappa=1$ and $\kappa=-1$. Here, $\theta_K$ and $\theta'_K$ represent different Kerr spectra.

In recent years, the detection and manipulation of antiferromagnetism is vital for spintronics, but technical challenging. The MOKE provides a powerful tool for detecting the electronic and magnetic properties of materials, which was first used to detect nonzero net magnetization in ferromagnets and ferrimagnets [48]. In antiferromagnets, though the time reversal ($T$) symmetry is broken, the combined operations of $T$ and crystal symmetry usually can preserve Kramers theorem, resulting in the absence of MOKE. Otherwise, MOKE will emerge. In 2015, it was first predicted that a large MOKE could emerge in noncollinear antiferromagnets $Mn_3X$ ($X$ = Rh, Ir, Pt) even though the net magnetization is zero [49]. Later, Zhou *et al*. theoretically studied the MOKE in noncollinear antiferromagnets $Mn_3XN$ ($X$ = Ga, Zn, Ag, and Ni) [50]. In 2016, Sivadas *et al*. predicted that the perpendicular electric field breaks the combined $TI$ symmetry ($I$ is spatial inversion), and thus induces MOKE in the bilayer $MnPSe_3$ with collinear antiferromagnetism [51]. In addition, MOKE was also predicted in the AFM heterostructures [34,52]. Experimentally, zero-field MOKE in noncollinear antiferromagnet was first reported in $Mn_3Sn$ bulk at room temperature [53], and then in $Mn_3Ge$ bulk [54]. These theoretical predictions and experimental observations confirm that MOKE can occur in antiferromagnets without net magnetization.

The noncollinear $W_3$-spin texture in $W_3Cl_8$ monolayer provides an opportunity to utilize the 5$d$ antiferromagnetism. It is fortunate that the trimeric Y-AFM order in the W$_3$Cl$_8$ monolayer can be detected by MOKE. It is noteworthy that this trimeric Y-AFM state is a noncollinear configuration composed by FM W$_3$ trimers, and thus this is actually a Y-AFM texture on the triangular lattice. It is different from previously reported MOKE-active noncollinear AFM states in those bulk materials, where the Y-AFM texture is on the Kagome lattice plane. Therefore, different symmetries appear in these two cases, and thereby different MOKEs can be observed. [49,50,54].

Here the time reversal symmetry is broken, and its combination with other crystal symmetry operation is also not a "good" symmetry. Thus the Kramers degeneracy is lifted, accompanied by a strong SOC from W's 5$d$ orbitals, which leads to a significant MOKE signal [49,55]. For example, in the +$P$ phase with $\kappa$=+1 and $\varphi$=0°, the MPG is 3$m'$ and the optical conductivity tensor (OCT) can be generally expressed as [56]:

$$\text{OCT }(3m') = \begin{pmatrix} \sigma_{xx} & \sigma_{xy} & 0 \\ -\sigma_{xy} & \sigma_{xx} & 0 \\ 0 & 0 & \sigma_{zz} \end{pmatrix}, \tag{4}$$

For those systems with threefold rotational symmetry, i.e. W$_3$Cl$_8$ monolayer, $\sigma_{xy} \neq 0$ means that the MOKE is activated. Our first-principles calculation indeed finds the maximum Kerr angle ($\theta_{\text{Kmax}}$) to be -0.357° under 1.5 eV incident light, as shown in Fig. 4(a), which is comparable to that of Mn$_3$$X$ ($X$=Rh, Ir, Pt) (0.2° ~ 0.6°) [49], and even larger than that of the monolayer ferromagnet CrI$_3$ (0.286°) [57]. The essential role of SOC in the occurrence of MOKE is also confirmed, namely the MOKE signal would vanish if the SOC is switched off. When the spatial inversion operation $I$ is performed on this system, the +$P$ phase with $\kappa$=+1 transforms into the -$P$ phase with $\kappa$=-1, but the Kerr spectrum remains unchanged, as displayed in Fig. 4(a). In contrast, when the time reversal operation $T$ is performed, the Kerr spectrum is completely opposite, as shown in Fig. 4(b).

The change of magnetic chirality can also tune the MOKE signal. For given +$P$ state and $\varphi = 0°$, both $\kappa$=1 and $\kappa$=-1 lead to the MPG 3$m'$, but their MOKE signals are not identical, as shown in Fig. 4(c). The absolute value of $\theta_{\text{Kmax}}$ is 0.357° (at 1.5 eV) for $\kappa$=+1, which is larger than 0.312° (at 1.2 eV) for $\kappa$=-1. This tendency is reasonable since a slightly larger $P$ exists in the $\kappa$=+1 case (due to the $P_\text{m}$). The effects of aforementioned operations, including $I$, $T$, and $C$ (the pure chirality switch between $\kappa$=1 and $\kappa$=-1), are summarized in Fig. 4(d).



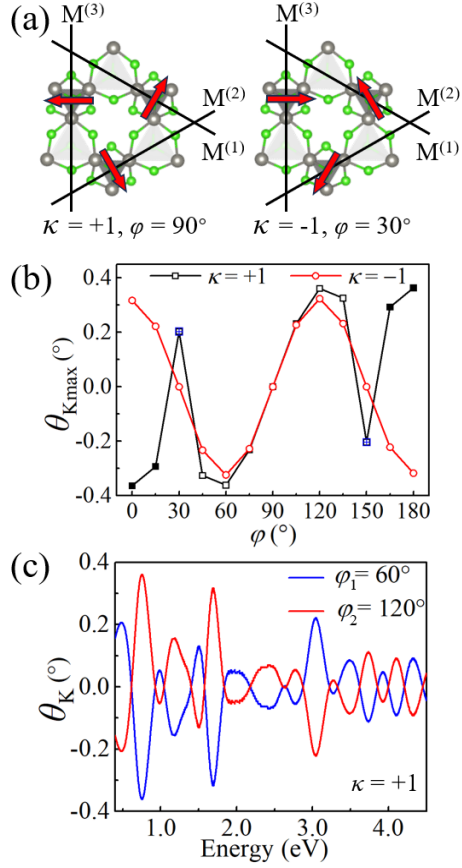

**FIG. 5.** (a) The magnetic configurations in the +P phase without MOKE. Here, $M^{(1)}$, $M^{(2)}$, $M^{(3)}$ represent the three mirror planes, and the red arrows represent the $W_3$-spins. (b) The maximum Kerr angle $\theta_{Kmax}$ as a function of $\varphi$ for $\kappa=\pm1$ in the same +P phase. The data of $\kappa=-1$ are obtained under 1.2 eV incident light. The data of $\kappa=+1$ marked by the hollow squares, solid squares and hollow squares with cross center are obtained under 0.8, 1.5, and 0.5 eV incident light respectively. (c) The Kerr spectra with $\varphi_1+\varphi_2=180°$ at $\kappa=+1$.

In addition, the MOKE signal of $W_3Cl_8$ monolayer is greatly influenced by the details of magnetic texture. Table I lists the $\varphi$-dependent MPG in the +P state for both $\kappa=1$ and -1. During the rotation of $\varphi$, new MPGs emerge, i.e. 3 and 3m. Their OCTs are [56]:

$$\text{OCT}(3) = \begin{pmatrix} \sigma_{xx} & \sigma_{xy} & 0 \\ -\sigma_{xy} & \sigma_{xx} & 0 \\ 0 & 0 & \sigma_{zz} \end{pmatrix}, \tag{5}$$

$$\text{OCT}(3m) = \begin{pmatrix} \sigma_{xx} & 0 & 0 \\ 0 & \sigma_{xx} & 0 \\ 0 & 0 & \sigma_{zz} \end{pmatrix}. \tag{6}$$

According to Eq. 1, in the 3$m$ case, $\sigma_{xy}$=0 and thus no MOKE exists. While for the MPG 3, nonzero $\sigma_{xy}$ means an observable MOKE, as listed in Table I. When $\varphi$=90° for $\kappa$=+1 or $\varphi$=30°, 90°, 150° for $\kappa$=-1, there is at least one of the three W$_3$-spins perpendicular to the corresponding mirror planes (M$^{(1)}$, M$^{(2)}$, M$^{(3)}$) of the structure, as illustrated in Fig. 5(a). The Kramers theorem is recovered by the combined $TM$ ($M$: the mirror operation) symmetry, and therefore the MOKE signal vanishes. Otherwise, $TM$ symmetry does not exist and the MOKE signal can be detected.

The nonzero Kerr spectrum exhibits a sensitive dependence on the detailed magnetic order. The $\varphi$ dependence of $\theta_{Kmax}$ for $\kappa$=±1 is summarized in Fig. 5(b). The absolute value of $\theta_{Kmax}$ reaches the maximum at $\varphi$=0°, 60°, 120°, 180°. Corresponding to the variation of magnetic group, different periodic behaviors are revealed for different chiralities, namely $\theta_{Kmax}$ as a function of $\varphi$ displays a period of 360° for $\kappa$=+1, but a period of 120° for $\kappa$=-1. In addition, for two trimeric Y-AFM states with different $\varphi$ (e.g. $\varphi_1$ and $\varphi_2$), if $\varphi_1+\varphi_2$=180°, opposite Kerr spectra appear as plotted in Fig. 5(c). The $\varphi$ modulation of the Kerr spectrum provides an efficient approach, in addition to the time reversal operation, to realize the flipping of the Kerr spectrum. Conversely, the magnetic-order dependent MOKE also provides an effective method to detect complex noncollinear AFM orders.

**TABLE I.** The magnetic point group (MPG) and $\sigma_{xy}$ for W$_3$Cl$_8$ monolayer in the same +$P$ phase for different magnetic states with $\kappa$=±1 and $\varphi$: 0°~180°.

|  | $\kappa$ | azimuthal angle $\varphi$ (°) | | | | | | |
|---|---|---|---|---|---|---|---|---|
|  |  | 0/180 | 15/165 | 30/150 | 45/135 | 60/120 | 75/105 | 90 |
| MPG | +1 | 3$m'$ | 3 | 3 | 3 | 3 | 3 | 3$m$ |
|  | -1 | 3$m'$ | 3 | 3$m$ | 3 | 3$m'$ | 3 | 3$m$ |
| Nonzero $\sigma_{xy}$ | +1 | yes | yes | yes | yes | yes | yes | no |
|  | -1 | yes | yes | no | yes | yes | yes | no |

**SUMMARY**

In conclusion, by using first-principles calculations, we predicted a stable multiferroic W$_3$Cl$_8$ monolayer with a breathing Kagome lattice. The reduced dimension and the fractional valence activate 5$d$ magnetism of tungsten, forming the molecular-orbital magnetic moment of W$_3$ trimers. The strong easy-plane magnetic anisotropy and the dominant AFM couplings lead to the noncollinear Y-AFM state composed by FM W$_3$ trimers as the magnetic ground state. It is revealed that an intrinsic MOKE with sizable Kerr angle appears in this AFM monolayer. Since the Y-AFM order of W$_3$-spins on triangular lattice may have different chiralities and diverse morphologies, corresponding to different symmetries, rich MOKEs are observed strongly dependent on the detailed magnetic order. Moreover, the evolutions of Kerr spectra with spin orientation are distinguishable for different chiralities, suggesting that MOKE can be used to detect the details of AFM orders. Meanwhile, the $W_3$ trimerization and the breathing distortion of Kagome lattice induce ion displacements, and thus result in a spontaneous ferroelectric polarization. This ferroelectric polarization can be reversed by the breathing process of the Kagome network, together with the displacement of magnetic trimers. Our study opens a route to realize more $5d$ magnetism/ferroelectricity/multiferroicity of tungsten-based compounds, especially in the low-dimensional forms, which may provide an excellent platform for investigating magneto-optical phenomena and exploring new technological applications.

## ACKNOWLEDGEMENTS

We thank Zhong Shen and Dr. Jun-Jie Zhang of Southeast University for fruitful discussions. This work is supported by Natural Science Foundation of Jiangsu Province (Grant No. BK20221451), National Natural Science Foundation of China (Grant Nos. 12325401, 12274069, 12274070). Most calculations were done on the Big Data Computing Center of Southeast University.

electronic structures with SOC considered; Note 3. The calculation of magnetic orders.